\documentclass[twocolumn]{aastex631x}
\begin{document}

\title{Rapid Evolution of the White Dwarf Pulsar AR Scorpii}

\correspondingauthor{Peter Garnavich}
\email{pgarnavi@nd.edu}

\author[0000-0003-4069-2817]{Peter Garnavich}
\affiliation{Department of Physics and Astronomy, University of Notre Dame, Notre Dame, IN 46556, USA}

\author{Stephen B. Potter}
\affiliation{South African Astronomical Observatory, PO Box 9, Observatory 7935, Cape Town, South Africa}
\affiliation{Department of Physics, University of Johannesburg, PO Box 524, 2006 Auckland Park, Johannesburg, South Africa}

\author{David A. H. Buckley}
\affiliation{South African Astronomical Observatory, PO Box 9, Observatory 7935, Cape Town, South Africa}
\affiliation{Department of Astronomy, University of Cape Town, Private Bag X3, Rondebosch 7701, South Africa}
\affiliation{Department of Physics, University of the Free State, PO Box 339, Bloemfontein 9300, South Africa}

\author[0000-0003-0397-4655]{Anke van Dyk}
\affiliation{South African Astronomical Observatory, PO Box 9, Observatory 7935, Cape Town, South Africa}
\affiliation{Department of Astronomy, University of Cape Town, Private Bag X3, Rondebosch 7701, South Africa}

\author{Daniel Egbo}
\affiliation{South African Astronomical Observatory, PO Box 9, Observatory 7935, Cape Town, South Africa}
\affiliation{Department of Astronomy, University of Cape Town, Private Bag X3, Rondebosch 7701, South Africa}

\author[0000-0001-7746-5795]{Colin Littlefield}
\affiliation{Bay Area Environmental Research Institute, Moffett Field, CA 94035 USA}
\affiliation{Department of Physics and Astronomy, University of Notre Dame, Notre Dame, IN 46556, USA}

\author[0009-0008-6389-9398]{Anousha Greiveldinger}
\affiliation{Department of Physics and Astronomy, University of Notre Dame, Notre Dame, IN 46556, USA}

\begin{abstract}

Analysis of AR~Sco optical light curves spanning nine years show a secular change in the relative amplitudes of the beat pulse pairs generated by the two magnetic poles of its rotating white dwarf. Recent photometry now shows that the primary and secondary beat pulses have similar amplitudes, while in 2015 the primary pulse was approximately twice that of the secondary peak. The equalization in the beat pulse amplitudes is also seen in the linearly polarized flux. This rapid evolution is consistent with precession of the white dwarf spin axis. The observations imply that the pulse amplitudes cycle over a period of $\gtrsim 40$~yrs, but that the upper limit is currently poorly constrained. If precession is the mechanism driving the evolution, then over the next 10 years the ratio of the beat pulse amplitudes will reach a maximum followed by a return to asymmetric beat pulses.

\end{abstract}


\section{Introduction} 

AR Scorpii (AR~Sco hereafter) is one of the most intriguing interacting binary stars known. It has been called a white dwarf ``pulsar'' \citep{buckley} because its bright, polarized, synchrotron flashes appear to be powered by the spin-down energy of its degenerate primary stellar component \citep{marsh, stiller18, gaibor20, pelisoli22}. The binary consists of a low-mass red dwarf star and a rapidly spinning magnetized white dwarf (WD) orbiting over a 3.56-hour period. The WD spins ($\omega$) with a period of $1.95$~min \citep{marsh} and appears to generate two pulses per rotation, likely from two magnetic poles. In the \citet{potter18} model, the synchrotron emission is modulated by the angle between the magnetic poles of the spinning WD and the secondary star. The emission is enhanced just after a magnetic pole sweeps past the red dwarf, leading to variations seen at the spin frequency plus strong pulses at the beat frequency, $\omega -\Omega$, where $\Omega$ is the orbital frequency. Because pulses are generated from both poles, power is also seen at the first harmonic of the beat frequency (aka the double-beat: {2($\omega-\Omega$)}).

Light curves of AR~Sco obtained near its discovery clearly showed that the beat pulse pairs alternated in strength with the brighter pole being about twice the amplitude of the opposite pole \citep{marsh}. This alternating strength was seen in the linearly polarized flux as well \citep{potter18}. However, in a Fourier component analysis, \citet{pelisoli22} noted that the quality of their model fit to the observed light curves was decreasing over time. They attributed this to stochastic changes in the relative strength of the pulse pairs. \citet{takata21} noted that the X-ray beat pulses were single peaked in 2016 but seen as double-peaked in 2020. Here, we analyze nine years of AR~Sco optical light curves to test for any systematic evolution in the beat pulse strengths.

\begin{deluxetable}{lccccc}
\centering
\tablecaption{Time-Series Photometry \label{photometry}}
\tablehead{ 
\multicolumn{6}{c}{Total Flux} \\
\hline
\colhead{Year} & \colhead{Start}  & \colhead{Cadence} & \colhead{Length} & \colhead{Telescope} & \colhead{$R_{beat}^{a}$}  \\
\colhead{} & \colhead{(MJD)} &   \colhead{(sec)}  & \colhead{(hours)} & \colhead{} & \colhead{}
}
\startdata
2015 & 57197.9  & 1.3   &  2.8  &  WHT  & 0.42$\pm0.03$ \\
2016 & 57463.0  &  1.0  &  2.3  &  SAAO & 0.53$\pm0.03$ \\
2016 & 57536.4  &  1.0  &  6.8  &  SAAO & 0.49$\pm0.03$ \\
2017 & 57891.3  &  2.0  &  4.0  &  SAAO & 0.37$\pm0.05$ \\
2018 & 58200.5  &  1.0  &  4.0  &  SAAO & 0.58$\pm0.03$ \\
2019 & 58613.4  &  1.0  &  5.0  &  SAAO & 0.66$\pm0.03$ \\
2020 & 59016.1  &  3.3  &  4.6  &  SLKT & 0.69$\pm0.04$ \\
2020 & 59037.1  &  2.5  &  4.5  &  SLKT & 0.62$\pm0.03$ \\
2021 & 59322.3  &  3.0  &  4.6  &  SLKT & 0.65$\pm0.03$ \\
2021 & 59342.2  &  2.6  &  4.6  &  SLKT & 0.77$\pm0.04$ \\
2022 & 59751.1  &  3.6  &  4.1  &  SLKT & 0.79$\pm0.03$ \\
2023 & 60090.2  &  3.9  &  4.2  &  SLKT & 0.79$\pm0.04$ \\
2023 & 60091.3  &  2.0  &  4.5  &  SAAO & 0.85$\pm0.03$ \\
2023 & 60092.4  &  2.4  &  4.9  &  SAAO & 0.82$\pm0.03$ \\
\hline
\multicolumn{6}{c}{Linearly Polarized Flux} \\
\hline
2016 & 57463.0  &  1.0  &  2.3  &  SAAO & 0.52$\pm0.03$ \\
2023 & 60091.3  &  2.0  &  4.5  &  SAAO & 0.81$\pm0.03$ \\
\enddata
\tablenotetext{a}{Beat/Double-Beat amplitude ratio calculated from Equation~1.}
\end{deluxetable}

\vspace{0.2cm}
\section{Data}

We analyze light curves studied by \citet{gaibor20}. In addition, we have obtained rapid-cadence light curves of AR~Sco using the Sarah L. Krizmanich Telescope (SLKT) from 2020 through 2023 and a light curves in 2023 from the South African Astronomical Observatory using the high-speed photomultiplier instrument (HIPPO, \citet{potter08}). We analyze both the total flux and polarized flux for two epochs obtained in 2016 and 2023 using the HIPPO instrument.  Only light curves covering a substantial fraction of a binary orbit are included in this study. Properties of the photometric time-series datasets are listed in Table~1.

\begin{figure}
\begin{center}
\includegraphics[scale=0.36,angle=0]{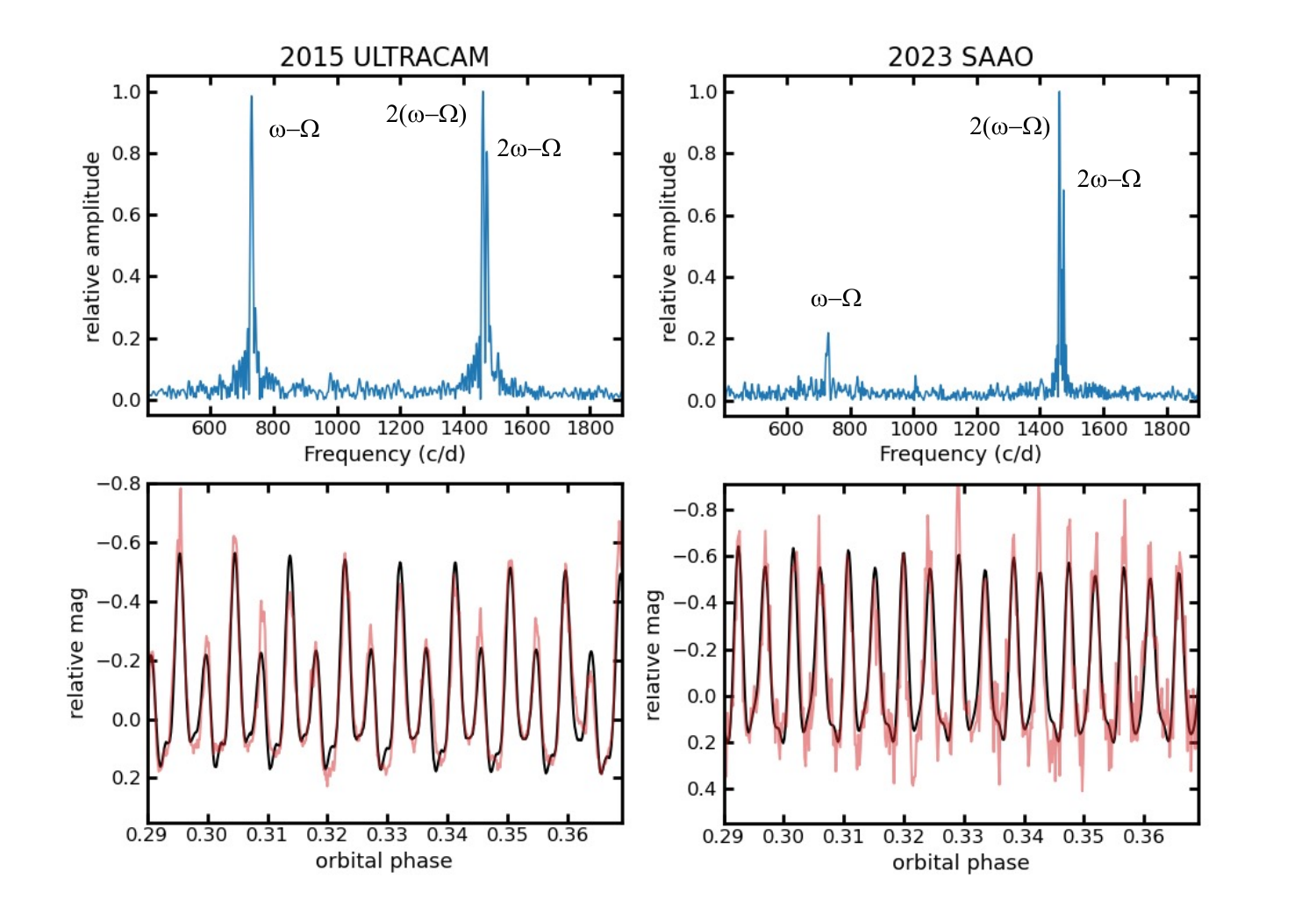}
\caption{{\bf Left Column:} The L-S periodogram (top) and section of the light curve (bottom) from 2015. The periodogram displays nearly equal amplitude peaks of the beat and double beat. This is seen in the light curve as the primary peak being twice the amplitude of the secondary one. The Fourier model is plotted as a solid black line and the data in faint red.  {\bf Right Column:} The L-S periodogram (top) and light curve (bottom) from 2023. The periodogram shows a very weak beat amplitude and a strong double beat peak. This is seen in the light curve as nearly equal amplitude beat pairs.   \label{fig1}}
\end{center}
\end{figure}

For each photometric time-series, we constructed Lomb-Scargle periodograms \citep[L-S hereafter,][]{lomb76,scargle82} implemented using the \emph{LombScargle} package in \emph{astropy} \citep{astropy18}. For example, L-S periodograms of the ULTRAcam photometry taken in 2015 and the SAAO photometry from 2023 are displayed in Figure~\ref{fig1}. From the strongest peaks in the periodogram ($\omega-\Omega$, $2(\omega-\Omega)$, $2\omega-\Omega$, $3(\omega-\Omega)$, $4(\omega-\Omega)$, $2\omega$, $2(2\omega-\Omega)$), we reconstructed the 2015 and 2023 light curves and these are also shown in Figure~\ref{fig1}.  The properties of the optical light curves appear to have changed significantly between 2015 and 2023. In particular, the strength of the `secondary' beat pulse has increased relative to the main pulse over this period.  

To quantify the relative strengths of the beat pulses averaged over an orbit, we use the L-S periodogram peaks at the beat frequency ($A_{(\omega-\Omega)}$) and the double-beat frequency ($A'_{2(\omega-\Omega)}$), to define the ratio:
\begin{equation}
R_{beat}= {{A'_{2(\omega-\Omega)}}\over{A_{(\omega-\Omega)}+A'_{2(\omega-\Omega)}}}\;\; .
\end{equation}

When $R_{beat} \approx 0$, only a single beat pulse is detected over a beat period. Alternatively, when $R_{beat}$ approaches unity, the two beat pulses are similar in amplitude. Here, $A'_{2(\omega-\Omega)}= A_{2(\omega-\Omega}) - H_2\, A_{(\omega-\Omega)}$ is the amplitude of the double beat peak corrected for a contribution of the second harmonic of the beat frequency. The parameter $H_2$ is the ratio between the amplitudes of the beat frequency, and its second harmonic. Simulated light curves show that an $H_2=0.30$ accounts for this harmonic contribution\footnote{Fourier analysis of Gaussian pulses in flux give a closed form for the harmonic ratio of $H_2=exp(-6\pi^2\sigma^2/P^2)$, where $P$ is the time between pulses. However, analyzing light curves in magnitudes requires simulations to estimate $H_2$.} that only becomes significant when when $R_{beat} < 0.5$, that is, when the secondary pulse is weak.

The measured beat pulse ratios, $R_{beat}$, for 14 light curves are given in Table~1. Uncertainties on the ratio measurements were estimated by taking L-S periodograms of subsets of each light curve. The variance in $R_{beat}$ was then used to calculate the errorbars shown. 
We also estimated the beat pulse ratio in linearly polarized flux using photometric time-series obtained in 2016 and 2023 and these results are shown in Table~1.

\section{Discussion} 

\begin{figure}
\begin{center}
\includegraphics[scale=0.50,angle=0]{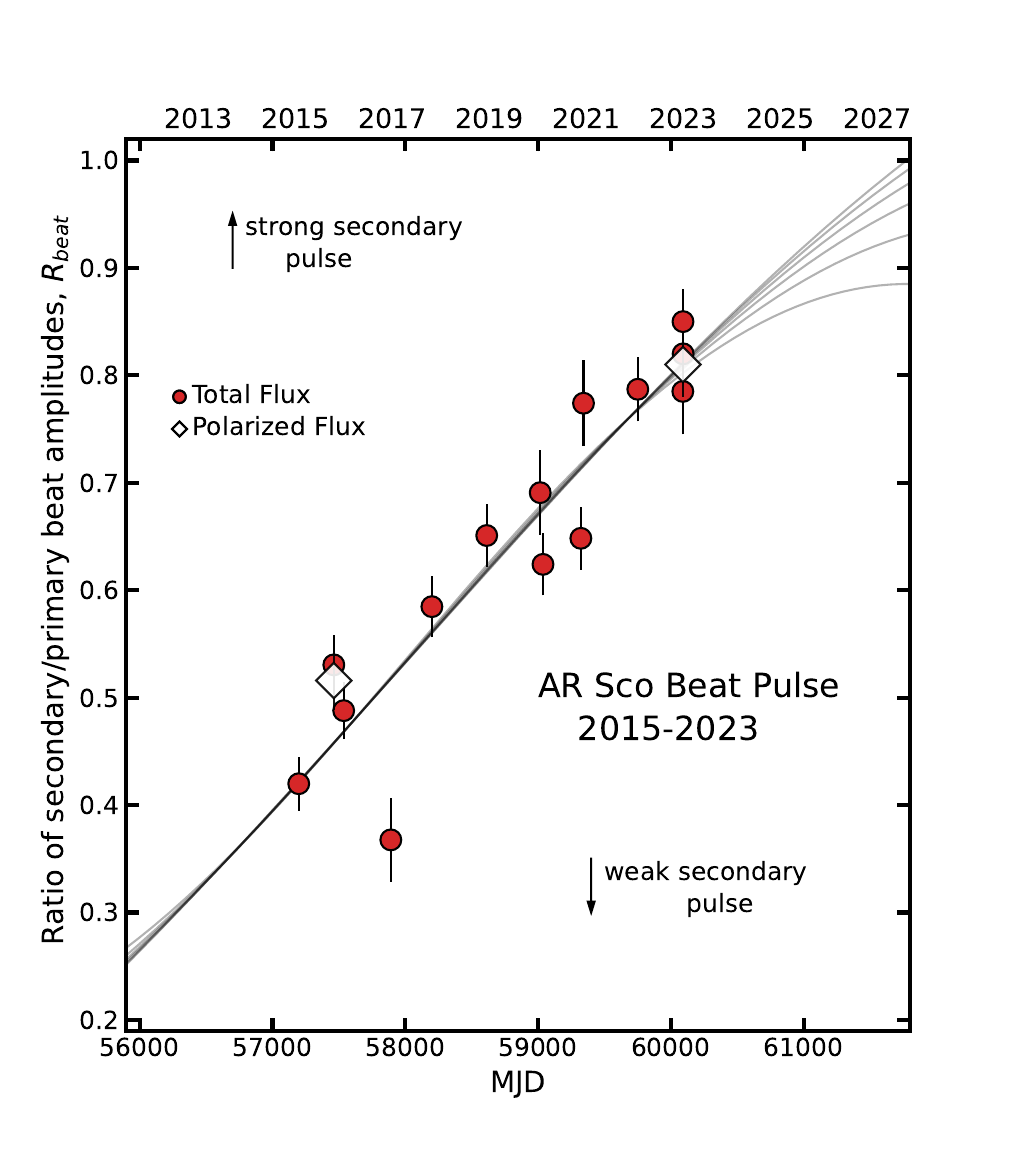}
\caption{The ratio of the double-beat to beat amplitudes measured for 14 light curves (solid circles). The $R_{beat}$ parameter measured from the linearly polarized flux in two epochs is displayed as open diamonds. The lines show a range cyclical models with periods between 30 and 100~yr. The $\chi^2$ parameter increases sharply for models with periods shorter than 40~years, but long periods are poorly constrained without further observations. \label{fig2}}
\end{center}
\end{figure}

Figure~\ref{fig2} displays the AR~Sco beat pulse ratio over time. While there is scatter in the ratio from night to night, the trend is for an increasing ratio consistent with the pulse pairs evolving to nearly equal strength. The $R_{beat}$ parameter for the linearly polarized flux was also seen to increase significantly between 2016 and 2023.

Interpreting the changes in pulse amplitude is difficult given that the source of the relativistic electrons has not been established. Electrons may be accelerated by direct interaction between the WD magnetic field and the field of the secondary star \citep[e.g.][]{takata17,garnavich19}. Slow changes in the secondary star's magnetic field could influence the WD field lines involved in trapping the emitting electrons.

\citet{katz17} proposed that the spin axis of the WD could be tilted relative to the orbital angular momentum vector, leading to a precession of the WD and its magnetic field configuration. Katz predicted that WD precession might impact the phase of the brightest point in the orbital modulation seen in AR~Sco. \citet{peterson} did not detect a shift in the orbital modulation curve using archival photometry\footnote{Also see \citet{littlefield17}}. However, the narrowness of the synchroton beams could provide a sensitive test of precessional motion if it results in relative changes in the viewing angles of the two poles.  \citet{katz17} predicted precession periods between 20 and 200 years primarily depending on the WD mass. 

\begin{figure}
\begin{center}
\includegraphics[scale=0.43,angle=0]{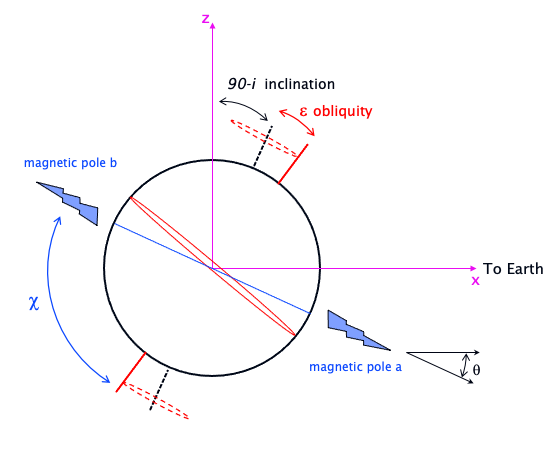}
\caption{The geometry of a precessing WD with synchrotron beams emitted from magnetic poles. The magnetic axis makes an angle $\chi$ relative to the spin axis (solid red lines). The binary orbital inclination is $i$, and the obliquity of the spin axis is $\epsilon$. The path of spin axis precession is indicated by dashed red circles. The angle between the direction to Earth and the peak of a synchrotron beam is $\theta$. For simplicity, the synchrotron beams are shown emanating from the WD, but they actually originate from the magnetic fields on the opposite side of the WD \citep[see][]{potter18}. \label{model}}
\end{center}
\end{figure}

\subsection{Precession Model}

To test if WD precession could generate the observed variation in pulse amplitudes, we constructed a simple model illustrated in Figure~\ref{model}. The model parameters include the binary inclination, $i$, the obliquity of the spin axis, $\epsilon$, and the angle between the magnetic dipole and the WD spin axis, $\chi$. From this geometric model, we calculate the minimum angle, $\theta$, between center of the synchrotron beams and the direction to the Earth and follow its variation over a precession cycle.

Based on the estimated binary mass ratio \citep{marsh}, and lack of detected eclipses the orbital inclination is $i\approx 75^\circ$ \citep{garnavich19}. That there are two pulses per WD spin suggests that the magnetic field axis is highly inclined to the spin axis \citep{geng16}, however, the symmetry of truly perpendicular rotator ($\chi > 80^\circ$) would severely limit the range of pulse ratio variations over a precession cycle. 

Following the beam model described in \citet{potter18}, we approximate the synchrotron pulse profile by a Gaussian function with $\sigma=40^\circ$. Thus, the intensity of the synchrotron beam will have fallen to half its peak value when viewed at an angle of $\theta \approx 47^\circ$. For simplicity, we assume that the peak intensity and profile distributions are identical for the two emitting poles.

\begin{figure}
\begin{center}
\includegraphics[scale=0.34,angle=0]{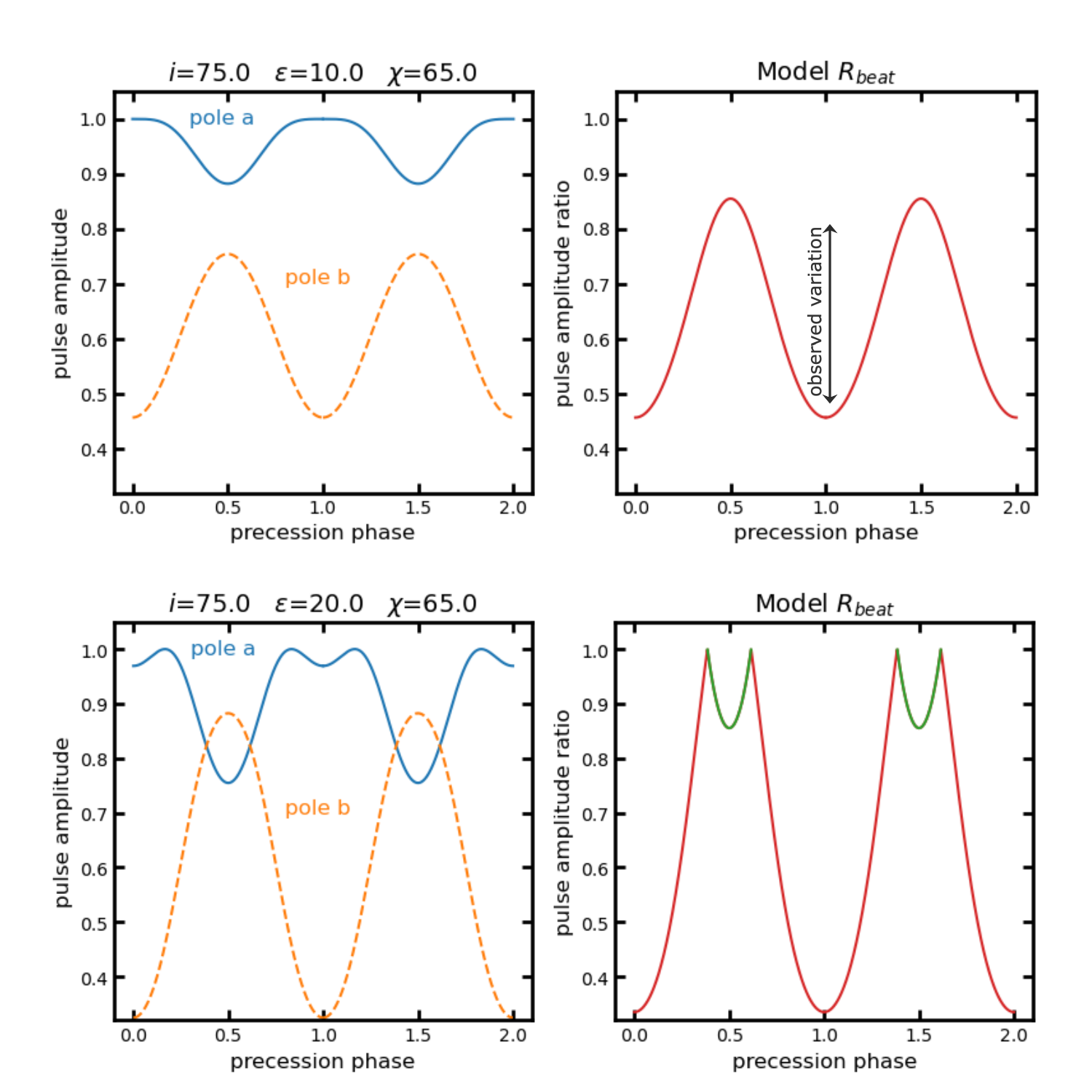}
\caption{Predictions for the $R_{beat}$ parameter as a function of precession phase for two sets of model parameters. {\bf Top:} The relative pulse amplitudes over two precession cycles for a WD obliquity of 10$^\circ$. The right panel shows that the resulting $R_{beat}$ parameter varies nearly linearly between 0.5 and 0.8 over half of a precession period. {\bf Bottom:} The same model except the obliquity has been increased to 20$^\circ$ and now exceeds the co-inclination of the orbit. In this case, the poles can switch in dominance and create sharp features in the $R_{beat}$ curve.  \label{model2}}
\end{center}
\end{figure}

We find that the observed change in $R_{beat}$ from 0.5 to 0.8 can be achieved by the precession model using reasonable values of the parameters (Figure~\ref{model2}). A WD obliquity near $\epsilon\approx 10^\circ$ is sufficient to reproduce the observations. Interestingly, when the obliquity exceeds the co-inclination ($90-i$), the secondary pulse strength can exceed the amplitude of the primary pulse, as shown in the lower panel of Figure~\ref{model2}. 

Besides the $R_{beat}$ ratio parameter, an additional observational constraint on precession models is a direct measurement of the amplitude changes of the pulses from each pole. As displayed in Figure~\ref{model2}, the amplitude of the pulse from the dominant pole varies by only 10\%\ over a quarter of the precession cycle for the first set of model parameters. In general, this small amplitude change for one pole results when the obliquity is low and $\chi\approx i$. For this geometry, changes in $\theta$ for one of the emitting poles remains small over a precession cycle, while the second pole creates most of the variation in the amplitude ratio.

The lower panels of Figure~\ref{fig1} show that the beat pulse amplitude of the dominant pole remained nearly the same between 2015 and 2023, and that most of the $R_{beat}$ evolution came from strengthening in the secondary pulse. We estimate that the amplitude of the primary pulse varied by no more than 0.1~mag over the nine years of observation. We therefore infer that $\chi\approx i$ for AR~Sco.

\subsection{Precession Period}

Despite the rapid variation observed in $R_{beat}$ over nine years, the apparent linear increase with time poorly constrains estimates of the precession period. Figure~\ref{fig2} displays sinusoidal functions with periods between 30 and 100~yr that have been fitted by minimizing the $\chi^2$ parameter. The $\chi^2$ parameter steeply increases for periods less than 40~yr, but $\chi^2$ is nearly constant for models with $P>40$~yr. A linear fit of the $R_{beat}$ ratio shows it rising by 0.05 per year, meaning that the pulses could reach parity as early as the year 2027 (MJD$\approx$61400). Sinusoidal extrapolations indicate that $R_{beat}$ will reach a maximum before the year 2029.

\section{Conclusion}

The AR~Sco beat pulse pairs have evolved from a strong asymmetry to become nearly equal in amplitude over a decade of observations. This evolution is supported by the changes noted in the X-ray beat pulse noted by \citet{takata21}. We also find that the beat pulses have evolved in linearly polarized flux and that their amplitudes are now nearly equal. This suggests that the evolution results from physical changes or viewing angle variations in the highly polarized synchrotron beams. 

The precession model suggests that in 2023 we were viewing the WD spin axis oriented so that the synchrotron beams are seen symmetrically, while in 2015, precession of the spin axis resulted in a tilt that favored our view of one of the magnetic poles. If precession is the origin of this evolution, then over the next 10 years the ratio of the beat pulse amplitudes, $R_{beat}$, will reach a maximum followed by a return to asymmetric beat pulses.


We dedicate this study to Tom Marsh. PMG thanks the Krizmanich family for their generous donations for the construction and support of the Sarah L. Krizmanich Telescope.

\end{document}